\documentclass[pra,aps,eqsecnum,amsmath,amssymb,showkeys]{revtex4}

\newcommand{\pen}{\openone}
\newcommand{\hh}{\mathcal{H}}
\newcommand{\lnp}{\mathcal{L}}
\newcommand{\lpp}{\mathcal{L}_{++}}
\newcommand{\lsp}{\mathcal{L}_{+}}
\newcommand{\qm}{\mathsf{Q}}
\newcommand{\rn}{\mathsf{R}}
\newcommand{\xm}{\mathsf{X}}
\newcommand{\zm}{\mathsf{Z}}
\newcommand{\um}{\mathsf{U}}
\newcommand{\vm}{\mathsf{V}}
\newcommand{\dm}{\mathsf{D}}
\newcommand{\wtv}{\widetilde{\mathsf{V}}}
\newcommand{\wtq}{\widetilde{\mathsf{Q}}}

\newcommand{\phim}{\mathrm{\Phi}}
\newcommand{\bro}{\boldsymbol{\rho}}
\newcommand{\bor}{\boldsymbol{\omega}}
\newcommand{\wro}{\widetilde{\boldsymbol{\rho}}}
\newcommand{\Tr}{{\mathrm{Tr}}}
\newcommand{\ri}{\mathrm{i}}
\newcommand{\hr}{{\mathrm{R}}}
\newcommand{\rh}{{\mathrm{H}}}
\newcommand{\ers}{{\mathrm{E}}_{\alpha}^{(s)}}
\begin{document}
\clearpage
\preprint{}

\title{Relations for certain symmetric norms and anti-norms before and after partial trace}
\author{Alexey E. Rastegin}
\affiliation{Department of Theoretical Physics, Irkutsk State University,
Gagarin Bv. 20, Irkutsk 664003, Russia}

\begin{abstract}
Changes of some unitarily invariant norms and anti-norms under the
operation of partial trace are examined. The norms considered form
a two-parametric family, including both the Ky Fan and Schatten
norms as particular cases. The obtained results concern operators
acting on the tensor product of two finite-dimensional Hilbert
spaces. For any such operator, we obtain upper bounds on norms of
its partial trace in terms of the corresponding dimensionality and
norms of this operator. Similar inequalities, but in the opposite
direction, are obtained for certain anti-norms of positive
matrices. Through the Stinespring representation, the results are
put in the context of trace-preserving completely positive maps.
We also derive inequalities between the unified entropies of a
composite quantum system and one of its subsystems, where
traced-out dimensionality is involved as well.
\end{abstract}


\maketitle

\pagenumbering{arabic}
\setcounter{page}{1}

\section{Introduction}\label{intr}

In quantum theory, the state of a subsystem of composite quantum
system is described by a reduced density operator. So, the
operation of partial trace is common in studying properties of
multipartite systems. Recent advances in quantum information have
lead to renewed interest to related issues. Extensions
\cite{ando,hansen06,petz86} of Lieb's concavity theorem \cite{lieb73a}
and additivity properties of the Wigner-Yanase entropy
\cite{CH10,hansen07,luo08} are very important examples. In
particular, we are interested in how used quantitative measures
may be changed by the partial trace. Many of frequently used
measures are determined in terms of matrix norms. The Schatten and
Ky Fan norms seem to be most important of them. The mentioned
question has been considered for the trace, Frobenius and spectral
norms in \cite{lidar1} and for the Ky Fan norms in \cite{rast103}.

The aim of the present work is to obtain similar results for more
unitarily invariant norms as well as for symmetric anti-norms. In
Section \ref{prelm}, we recall required facts about the operation
of partial trace. We consider the two-parameter family of norms,
which includes both the Ky Fan and Schatten ones. The results for
norms are proved in Section \ref{mres}. In Section \ref{antr}, we
obtain relations of discussed type for a class of symmetric
anti-norms of positive operators. Using the Stinespring
representation, many of these results are reformulated for norms
and anti-norms of the output of a quantum channel. In Section
\ref{inuen}, the presented bounds are used for deriving
inequalities that relate the quantum $(\alpha,s)$-entropies of a
composite system and one of its subsystems.

\section{Preliminaries}\label{prelm}

Let $\lnp(\hh)$ be the space of linear operators on
$m$-dimensional Hilbert space $\hh$. By $\lsp(\hh)$ and
$\lpp(\hh)$, we respectively denote the set of positive
semidefinite operators and the set of strictly positive ones. A
unitarily invariant norm, in signs $|||\centerdot|||$, is a norm
on square matrices that obeys $|||\qm|||=|||\um\qm\vm|||$ for all
$\qm\in\lnp(\hh)$ and for unitary $\um$, $\vm$ \cite{bhatia}. For
any $\qm\in\lnp(\hh)$, we define $|\qm|\in\lsp(\hh)$ as the
positive square root of $\qm^{\dagger}\qm$. The singular values
$\sigma_{j}(\qm)$ are defined as the eigenvalues of $|\qm|$
\cite{bhatia}. The Schatten and Ky Fan norms both form especially
important families of unitarily invariant norms. For each real
number $p\geq1$, the Schatten $p$-norm is defined as
\cite{bhatia,hjhon85}
\begin{equation}
\|\qm\|_{p}:=\Bigl(\Tr\bigl(|\qm|^{p}\bigr)\Bigr)^{1/p}=
\Bigl(\sum\nolimits_{j=1}^{m}\sigma_{j}(\qm)^{p}{\,}\Bigr)^{1/p}
\ . \label{schnd}
\end{equation}
This family includes the trace norm $\|\qm\|_{1}=\Tr|\qm|$ for
$p=1$, the Frobenius norm
$\|\qm\|_{2}=\sqrt{\Tr(\qm^{\dagger}\qm)}$ for $p=2$, and the
spectral norm
$\|\qm\|_{\infty}=\max\{\sigma_{j}(\qm):{\>}1\leq{j}\leq{m}\}$ for
$p=\infty$. The trace and spectral norms are widely
used in upper estimates of Fannes type \cite{alf04,AE05,AE11,rastjmp}.
Recall that Fannes' inequality gives an upper bound on a
change of the von Neumann entropy for an argument variation
\cite{fannes}. The right-hand side of (\ref{schnd}) is actually
the ordinary $p$-norm of the vector $\sigma(\qm)$. Here the vector
$p$-norm is defined for $p\in[1;\infty]$ as
\begin{equation}
\|x\|_{p}:=\left(\sum\nolimits_{j=1}^{m}|x_{j}|^{p}\right)^{1/p}
\ . \label{pnmd}
\end{equation}
The function (\ref{pnmd}) is an example of symmetric gauge
function. For each integer $k=1,\ldots,m$, the Ky Fan $k$-norm is
defined as
\begin{equation}
\|\qm\|_{(k)}:=\sum\nolimits_{j=1}^{k}\sigma_{j}(\qm)^{\downarrow}
\ , \label{kyfnd}
\end{equation}
where the non-increasing order
$\sigma_{1}(\qm)^{\downarrow}\geq\sigma_{2}(\qm)^{\downarrow}\geq\ldots\geq\sigma_{m}(\qm)^{\downarrow}$
is assumed. The $k$-norm is related to the symmetric gauge function
\begin{equation}
G_{(k)}(x):=\sum\nolimits_{j=1}^{k}|x_{j}|^{\downarrow}
\ . \label{gaugf}
\end{equation}
Note that the family (\ref{kyfnd}) includes both the spectral norm
$\|\qm\|_{(1)}\equiv\|\qm\|_{\infty}$ and trace one
$\|\qm\|_{(m)}\equiv\|\qm\|_{1}$.

It is known that every unitarily
invariant norm can be defined via the corresponding symmetric
gauge function (for details, see sect. IV.2 in \cite{bhatia} or
7.4 in \cite{hjhon85}). Let $G(x)$ be a symmetric gauge function.
Then the function
\begin{equation}
G^{(p)}(x):=\Bigl(G\bigl(|x|^{p}\bigr)\Bigr)^{1/p}
\ , \label{paugf}
\end{equation}
written for $p\geq1$, is symmetric gauge as well \cite{bhatia}.
Applying this to (\ref{gaugf}), we obtain the corresponding
symmetric gauge function and unitarily invariant norm. For integer
$k=1,\ldots,m$ and real $p\geq1$, these terms are defined by
\begin{align}
&G_{(k)}^{(p)}(x):=\Bigl(\sum\nolimits_{j=1}^{k}\bigl(|x_{j}|^{\downarrow}\bigr)^{p}{\,}\Bigr)^{1/p}
\ , \label{kpgf}\\
&\|\qm\|_{(k)}^{(p)}:=G_{(k)}^{(p)}\bigl(\sigma(\qm)\bigr)=
\Bigl\{\sum\nolimits_{j=1}^{k}\bigl(\sigma_{j}(\qm)^{\downarrow}\bigr)^{p}{\,}\Bigr\}^{1/p}
\ . \label{npgf}
\end{align}
This two-parametric family includes both the Ky Fan and Schatten
norms as particular cases. In the following, we will study changes
of unitarily invariant norms of the form (\ref{npgf}) under the
operation of partial trace.

Let $\hh_{A}$ and $\hh_{B}$ be finite-dimensional Hilbert spaces
of dimensionality ${\rm{dim}}(\hh_{A})=m$ and
${\rm{dim}}(\hh_{B})=n$. By $\bigl\{|e_{i}\rangle\bigr\}$ and
$\bigl\{|f_{j}\rangle\bigr\}$, where $0\leq{i}\leq{m}-1$ and
$0\leq{j}\leq{n}-1$, we respectively denote some orthonormal bases
in $\hh_{A}$ and $\hh_{B}$. Consider operators of a kind
$|a'\rangle\langle{a}''|\otimes|b'\rangle\langle{b}''|$ with any
vectors $|a'\rangle,|a''\rangle\in\hh_{A}$ and
$|b'\rangle,|b''\rangle\in\hh_{B}$. For such operators, the
partial trace over $\hh_{A}$ and the partial trace over $\hh_{B}$
are defined as
\begin{align}
 & \Tr_{A}\Bigl(|a'\rangle\langle{a}''|\otimes|b'\rangle\langle{b}''|\Bigr)
:=\langle{a}''|a'\rangle{\>}|b'\rangle\langle{b}''|
\ , \label{prtdfa}\\
 & \Tr_{B}\Bigl(|a'\rangle\langle{a}''|\otimes|b'\rangle\langle{b}''|\Bigr)
:=\langle{b}''|b'\rangle{\>}|a'\rangle\langle{a}''|
\ . \label{prtdfb}
\end{align}
The definition is completed by requiring that the partial trace be
linear in its input \cite{nielsen}. Such a definition is rather
physicist-friendly. For description of partial trace as a linear
map, see, e.g., sect. 5.4 in \cite{carlen09}. For arbitrary
operator $\wtq\in\lnp\bigl(\hh_{A}\otimes\hh_{B}\bigr)$, we can
write the expression
\begin{equation}
\wtq=\sum\nolimits_{i,j=0}^{m-1}|e_{i}\rangle\langle{e}_{j}|\otimes\qm_{ij}
\ . \label{wtdec}
\end{equation}
With respect to chosen bases, this operator is represented as the
$m$-by-$m$ block matrix $\bigl[\bigl[\qm_{ij}\bigr]\bigr]$, in
which the submatrices $\qm_{ij}$ are of size $n\times{n}$. Such
representation assumes the Kronecker product. Then the partial
traces of any matrix $\wtq$ are expressed as
\begin{equation}
\Tr_{A}\bigl(\wtq\bigr)=\sum\nolimits_{i=0}^{m-1}\qm_{ii}
\ , \qquad
\Tr_{B}\bigl(\wtq\bigr)=\bigl[\bigl[\Tr(\qm_{ij})\bigr]\bigr]
\ . \label{qrtdf}
\end{equation}
That is, tracing-out $\hh_{A}$ leads to the sum of $m$ diagonal
submatrices; tracing-out $\hh_{B}$ gives the $m$-by-$m$ matrix
such that each submatrix $\qm_{ij}$ has been replaced with its
trace.

Let $\xm_{B}$ and $\zm_{B}$ be generalized Pauli operators on
$\hh_{B}$. These operators act as
$\xm_{B}|f_{j}\rangle=|f_{j+1}\rangle$ and
$\zm_{B}|f_{j}\rangle=\exp({\ri}2\pi{j}/n)|f_{j}\rangle$
\cite{JR10}. The following relations can immediately be checked:
\begin{align}
& \frac{1}{n}{\>}\sum\nolimits_{j=0}^{n-1}\bigl(\pen_{A}\otimes\zm_{B}^{j}\bigr)
{\,}\wtq{\,}\bigl(\pen_{A}\otimes\zm_{B}^{-j}\bigl)=
\bigl[\bigl[\dm_{ij}\bigr]\bigr]
\ , \label{zrel}\\
&\sum\nolimits_{l=0}^{n-1}\bigl(\pen_{A}\otimes\xm_{B}^{l}\bigr)
\bigl[\bigl[\dm_{ij}\bigr]\bigr]\bigl(\pen_{A}\otimes\xm_{B}^{-l}\bigl)=
\bigl[\bigl[\Tr(\dm_{ij})\bigr]\bigr]\otimes\pen_{B}
\ . \label{xrel}
\end{align}
Here the diagonal matrix $\dm_{ij}$ is obtained from $\qm_{ij}$ by
replacing all its off-diagonal entries with zeros. Since
$\Tr(\dm_{ij})=\Tr(\qm_{ij})$, the set of $n^{2}$ unitary matrices
$\xm_{B}^{l}\zm_{B}^{j}$ (where $l,j=0,1,\ldots,n-1$) fulfills the
formula
\begin{equation}
\frac{1}{n}{\>}\sum_{l=0}^{n-1}\sum_{j=0}^{n-1}\bigl(\pen_{A}\otimes\xm_{B}^{l}\zm_{B}^{j}\bigr)
{\,}\wtq{\,}\bigl(\pen_{A}\otimes\xm_{B}^{l}\zm_{B}^{j}\bigl)^{\dagger}=\Tr_{B}\bigl(\wtq\bigr)\otimes\pen_{B}
\ . \label{xzrel}
\end{equation}
With relevant changes, similar relation can be posed for
$\pen_{A}\otimes\Tr_{A}\bigl(\wtq\bigr)$. The formula
(\ref{xzrel}) is used in proving the monotonicity of quantum
relative entropies under partial trace  \cite{JR10}. Hence, the
monotonicity under trace-preserving completely positive maps can
be derived (see \cite{hmp10,JR10} and references therein). The
partial trace is a primary example of such maps, which are often
called ''quantum channels'' \cite{nielsen}. In sect. 7.2 of
\cite{carlen09}, one presents convexity properties of some
functionals defined in terms of partial trace. We will use
(\ref{xzrel}) in studying changes of unitarily invariant norms
(\ref{npgf}) under partial trace.

The partial trace is involved into both the Stinespring
representation \cite{stins55} and Choi--Jamio{\l}kowski
representation \cite{choi75,jam72} of quantum channels. Suppose
that linear map $\phim:{\>}\lnp(\hh_{A})\rightarrow\lnp(\hh_{B})$
is completely positive. There exist a linear operator $\wtv$ from
$\hh_{A}$ to $\hh_{B}\otimes\hh_{C}$ such that \cite{watrous1}
\begin{equation}
\phim(\qm)=\Tr_{C}\bigl(\wtv\qm\wtv^{\dagger}\bigr)
\label{strp}
\end{equation}
for all $\qm\in\lnp(\hh_{A})$. The right-hand side of (\ref{strp})
is the Stinespring representation of $\phim$. The dimensionality of
$\hh_{C}$ coincides with the rank of the Choi matrix (for details, see
sect. 5.2 in \cite{watrous1}). In the literature, this matrix is
also known as dynamical matrix \cite{bengtsson}. Incidentally, its
rank gives the number of Kraus operators in the canonical Kraus
form \cite{bengtsson}. When completely positive map $\phim$
preserves the trace, there holds
\begin{equation}
\wtv^{\dagger}\wtv=\pen_{A}
\ . \label{strp1}
\end{equation}
That is, the $\wtv$ is an isometry. The Stinespring representation
(\ref{strp}) is closely related to environmental description of a
quantum channel. Using (\ref{strp}), we will obtain some relations
for norms and anti-norms of the output $\phim(\qm)$.

\section{Main results for norms}\label{mres}

In this section, we prove desired relations between
corresponding norms of operator
$\wtq\in\lnp\bigl(\hh_{A}\otimes\hh_{B}\bigr)$ itself and one of
its partial traces. The following statement takes place.

\newtheorem{t31}{Proposition}
\begin{t31}\label{prop1}
Let $\wtq\in\lnp\bigl(\hh_{A}\otimes\hh_{B}\bigr)$,
${\rm{dim}}(\hh_{A})=m$, ${\rm{dim}}(\hh_{B})=n$, and
$\qm_{A}=\Tr_{B}\bigl(\wtq\bigr)$. For all $k=1,\ldots,m$ and
$p\geq1$, there holds
\begin{equation}
\|\qm_{A}\|_{(k)}^{(p)}\leq{n}^{(p-1)/p}{\>}\|\wtq\|_{(kn)}^{(p)}
\ . \label{kpn1}
\end{equation}
\end{t31}

{\bf Proof.} Applying the triangle inequality and the homogeneity
of norms to (\ref{xzrel}), we first observe that
\begin{align}
\bigl\|\qm_{A}\otimes\pen_{B}\bigr\|_{(kn)}^{(p)}&\leq
\frac{1}{n}{\>}\sum_{l=0}^{n-1}\sum_{j=0}^{n-1}
\left\|\bigl(\pen_{A}\otimes\xm_{B}^{l}\zm_{B}^{j}\bigr){\,}
\wtq{\,}\bigl(\pen_{A}\otimes\xm_{B}^{l}\zm_{B}^{j}\bigl)^{\dagger}\right\|_{(kn)}^{(p)}
\nonumber\\
&=n{\,}\|\wtq\|_{(kn)}^{(p)}
\ , \label{kpn12}
\end{align}
since the norms (\ref{npgf}) are all unitarily invariant. We also
see that
$\bigl|\qm_{A}\otimes\pen_{B}\bigr|=|\qm_{A}|\otimes\pen_{B}$ for
any $\qm_{A}\in\lnp(\hh_{A})$. Repeating the spectrum of
$|\qm_{A}|$ by $n$ times, one then obtains the spectrum of
$\bigl|\qm_{A}\otimes\pen_{B}\bigr|$. Hence we write
\begin{equation}
\bigl\|\qm_{A}\otimes\pen_{B}\bigr\|_{(kn)}^{(p)}=
\Bigl\{n\sum\nolimits_{j=1}^{k}\bigl(\sigma_{j}(\qm_{A})^{\downarrow}\bigr)^{p}{\,}\Bigr\}^{1/p}
=n^{1/p}{\,}\|\qm_{A}\|_{(k)}^{(p)}
\ . \label{kpn13}
\end{equation}
Combining (\ref{kpn12}) and (\ref{kpn13}) gives the inequality
\begin{equation}
n^{1/p}{\,}\|\qm_{A}\|_{(k)}^{(p)}\leq{n}{\,}\|\wtq\|_{(kn)}^{(p)}
\ , \label{kpn14}
\end{equation}
which is equivalent to the claim (\ref{kpn1}). $\blacksquare$

The inequality (\ref{kpn1}) provides an upper bound on the norms
(\ref{npgf}) of partial trace $\qm_{A}=\Tr_{B}\bigl(\wtq\bigr)$ in
terms of similar norms of $\wtq$ and traced-out dimensionality.
This inequality is sharp in the following sense. It is saturated
for all $k=1,\ldots,m$ and $p\geq1$, when the operator $\wtq$ is a
multiple of $\rn_{A}\otimes\pen_{B}$ with some
$\rn_{A}\in\lnp(\hh_{A})$. We then write
\begin{equation}
\wtq=c{\,}\rn_{A}\otimes\pen_{B}
\ , \qquad
\qm_{A}=\Tr_{B}\bigl(\wtq\bigr)=c{\,}n{\,}\rn_{A}
\ , \label{wrqa}
\end{equation}
where $c$ denotes a complex number. Substituting these expressions
into (\ref{kpn12}), we have actually arrived at the equality.
Setting $k=m$ in (\ref{kpn1}), we obtain relations between the
Schatten norms in the form
\begin{equation}
\|\qm_{A}\|_{p}\leq{n}^{(p-1)/p}{\>}\|\wtq\|_{p}
\ , \label{spn1}
\end{equation}
where $p\geq1$. Taking $p=1$, $p=2$, and $p\to\infty$, we further
obtain the inequalities
\begin{equation}
\|\qm_{A}\|_{1}\leq\|\wtq\|_{1}
\ , \qquad
\|\qm_{A}\|_{2}\leq\sqrt{n}{\>}\|\wtq\|_{2}
\ , \qquad
\|\qm_{A}\|_{\infty}\leq{n}{\>}\|\wtq\|_{\infty}
\ , \label{tfsn}
\end{equation}
for the trace, Frobenius and spectral norms, respectively. The
formulas (\ref{tfsn}) were obtained in \cite{lidar1} on base of a
certain integral representation for partial trace. Following from
Shur's lemma \cite{DPS04}, this representation uses an
integration with respect to the normalized Haar measure. We
have derived (\ref{tfsn}) with use of the elementary
representation (\ref{xzrel}). Choosing $p=1$ in (\ref{kpn1}), we have
the inequality in terms of the Ky Fan norms, namely
\begin{equation}
\|\qm_{A}\|_{(k)}\leq\|\wtq\|_{(kn)}
\ . \label{kpk1}
\end{equation}
For $k=m$, the relation (\ref{kpk1}) leads to the first inequality
of (\ref{tfsn}) with the trace norm. For $k=1$, we obtain
\begin{equation}
\|\qm_{A}\|_{\infty}\leq\|\wtq\|_{(n)}
\ . \label{kpk2}
\end{equation}
This is stronger than the third inequality of (\ref{tfsn}) (except
when multiplicity of the largest singular value of $\wtq$ is not
less than ${\rm{dim}}(\hh_{B})=n$). Note that the relation
(\ref{kpk1}) was previously obtained in \cite{rast103} with use of
the Ky Fan maximum principle \cite{kyfan}.

Combining the relation (\ref{kpn1}) with the Stinespring
representation (\ref{strp}), we can obtain upper bounds on
corresponding norms of the output of a quantum channel. Here,
we immediately obtain
\begin{equation}
\|\phim(\qm)\|_{(k)}^{(p)}\leq{d}^{(p-1)/p}{\>}\bigl\|\wtv\qm\wtv^{\dagger}\bigr\|_{(kd)}^{(p)}
\ , \label{strp2}
\end{equation}
where $d={\rm{dim}}(\hh_{C})$. For trace-preserving completely
positive map $\phim$, the operator $\wtv$ in (\ref{strp}) is an
isometry. It then follows from (\ref{strp1}) that
\begin{equation}
\bigl|\wtv\qm\wtv^{\dagger}\bigr|=\wtv|\qm|\wtv^{\dagger}
\ . \label{strp3}
\end{equation}
By the spectral decomposition of $|\qm|$, this formula shows that
the matrices $\wtv\qm\wtv^{\dagger}$ and $\qm$ have the same
non-zero singular values. Combining this fact with (\ref{strp2})
leads to the following statement.

\newtheorem{ct1}{Corollary}
\begin{ct1}\label{crl1}
Let $\phim:{\>}\lnp(\hh_{A})\rightarrow\lnp(\hh_{B})$ be
trace-preserving completely positive map with
${\rm{dim}}(\hh_{B})=n$, and let $\qm\in\lnp(\hh_{A})$. For all
$k=1,\ldots,n$ and $p\geq1$, there holds
\begin{equation}
\|\phim(\qm)\|_{(k)}^{(p)}\leq{d}^{(p-1)/p}{\>}\|\qm\|_{(kd)}^{(p)}
\ , \label{stct1}
\end{equation}
where $d$ is the dimensionality of $\hh_{C}$ in the Stinespring
representation (\ref{strp}).
\end{ct1}

For any input $\qm\in\lnp(\hh_{A})$, the inequality (\ref{stct1})
gives an upper bound on the norms (\ref{npgf}) of channel output
$\phim(\qm)$. In particular, the Schatten $p$-norm obeys
\begin{equation}
\|\phim(\qm)\|_{p}\leq{d}^{(p-1)/p}{\>}\|\qm\|_{p}
\ , \label{stctp}
\end{equation}
where real $p\geq1$. We shall now prove another statement, which
can be applied in combination with (\ref{kpn1}).

\newtheorem{t32}[t31]{Proposition}
\begin{t32}\label{prop2}
Let $\rn\in\lnp(\hh)$ and ${\rm{dim}}(\hh)=m$. For all
$k=1,\ldots,m$ and $p,q\geq1$, there holds
\begin{equation}
\|\rn\|_{(k)}^{(p)}\leq{k}^{(q-1)/(pq)}{\>}\|\rn\|_{(k)}^{(pq)}
\ , \label{tpn2}
\end{equation}
with equality if and only if multiplicity of the largest singular
value of $\rn$ is not less than $k$.
\end{t32}

{\bf Proof.} In line with the H\"{o}lder inequality for vector
norms of $k$-tuples $x$ and $y$, we have \cite{hardy}
\begin{equation}
|\langle{x}{\,},{y}\rangle|
\leq\|x\|_{q}{\>}\|y\|_{r}
\ . \label{hlin}
\end{equation}
Here the conjugate indices $q$ and $r$ obey $1/q+1/r=1$. Let us
put $x_{j}=\bigl(\sigma_{j}(\rn)^{\downarrow}\bigr)^{p}$ and
$y_{j}=1$ for all $1\leq{j}\leq{k}$. It then follows from
(\ref{hlin}) that
\begin{equation}
\sum\nolimits_{j=1}^{k}\bigl(\sigma_{j}(\rn)^{\downarrow}\bigr)^{p}\leq
\Bigl\{\sum\nolimits_{j=1}^{k}\bigl(\sigma_{j}(\rn)^{\downarrow}\bigr)^{pq}{\,}\Bigr\}^{1/q}k^{1-1/q}
\ . \label{gear}
\end{equation}
Raising this to the power $1/p>0$, we obtain (\ref{tpn2}). The
inequality (\ref{hlin}) is saturated if and only if $x_{j}^{q}$
and $y_{j}^{r}$ are proportional (see theorem 13 in \cite{hardy}).
So the equality in (\ref{gear}) is equivalent to that
$\sigma_{1}(\rn)^{\downarrow}=\sigma_{2}(\rn)^{\downarrow}=\ldots=\sigma_{k}(\rn)^{\downarrow}$,
i.e. multiplicity of the largest singular value of $\rn$ is not
less than $k$. $\blacksquare$

Proposition \ref{prop2} is an extension of the statement proved
for the Schatten norms in \cite{rast11a}. Combining (\ref{kpn1})
and (\ref{tpn2}) leads to the inequality
\begin{equation}
\|\qm_{A}\|_{(k)}^{(p)}\leq\left[k^{q-1}{\,}n^{pq-1}\right]^{1/(pq)}\|\wtq\|_{(kn)}^{(pq)}
\ , \label{cpn1}
\end{equation}
in which $k=1,\ldots,m$ and $p,q\geq1$. In some respects, these
results are complementary to the inequalities given in section IV
of \cite{AE05}. Upper bounds on unitarily invariant norms of a
traceless Hermitian operator in terms of the trace norm are
provided therein. The relation (\ref{tpn2}) holds for all
operators and gives an upper bound on the norm (\ref{npgf}) in
terms of similar norms with the same $k$ and larger real parameter
$pq$. The results of this section characterize relations between
some unitarily invariant norms of an operator and one of its
partial traces. In the next section, we obtain relations of such a
kind for symmetric anti-norms of positive operators.

\section{Relations for symmetric anti-norms}\label{antr}

In the recent papers \cite{bh10a,bh10b}, Bourin and Hiai examined
symmetric anti-norms of positive operators. They form a class of
functionals containing the right-hand side of (\ref{schnd}) for
$p\in(0;1)$ and, with strictly positive matrices, for $p<0$. For
arbitrary $\qm\in\lsp(\hh)$, we consider a functional
\begin{equation}
\qm\mapsto\|\qm\|_{!}
\ , \label{funex}
\end{equation}
taking values on $[0;\infty)$. If this functional enjoys the
homogeneity, the symmetry
$\|\qm\|_{!}=\|\um\qm\um^{\dagger}\|_{!}$ for all unitary $\um$,
and the superadditivity
\begin{equation}
\|\qm+\rn\|_{!}\geq\|\qm\|_{!}+\|\rn\|_{!}
\ , \label{sups}
\end{equation}
we call it a {\it symmetric anti-norm} \cite{bh10a}. In general,
anti-norms may vanish for non-zero operators. An important class
of symmetric anti-norms is formed by the Ky Fan ones. For integer
$k=1,\ldots,m$, where $m={\rm{dim}}(\hh)$, the Ky Fan
$k$-anti-norm is defined as the sum of the $k$ smallest
eigenvalues of $\qm\in\lsp(\hh)$, namely
\begin{equation}
\|\qm\|_{\{k\}}:=\sum\nolimits_{j=1}^{k}\lambda_{j}(\qm)^{\uparrow}=\Tr(\qm)-\|\qm\|_{(m-k)}
\ . \label{kyfad}
\end{equation}
Here the non-decreasing order
$\lambda_{1}(\qm)^{\uparrow}\leq\lambda_{2}(\qm)^{\uparrow}\leq\ldots\leq\lambda_{m}(\qm)^{\uparrow}$
is assumed. Combining (\ref{kyfad}) with
$\Tr(\qm+\rn)=\Tr(\qm)+\Tr(\rn)$ and the triangle inequality for
the norms
\begin{equation}
\|\qm+\rn\|_{(m-k)}\leq\|\qm\|_{(m-k)}+\|\rn\|_{(m-k)}
\ , \label{ksap}
\end{equation}
we obtain the superadditive property, that is
\begin{equation}
\|\qm+\rn\|_{\{k\}}\geq\|\qm\|_{\{k\}}+\|\rn\|_{\{k\}}
\ . \label{sapk}
\end{equation}
Note that the formula (\ref{ksap}) directly follows from the Ky
Fan maximum principle \cite{kyfan}. In paper \cite{uhlmann00},
Uhlmann examined so-called partial fidelities. So, the $k$-th
partial fidelity of density operators $\bro$ and $\bor$ is
actually the Ky Fan $\{m-k\}$-anti-norm of
$\left|\sqrt{\bro}\sqrt{\bor}\right|$. For $p\in(0;1)$, the
Schatten $p$-anti-norm is defined as \cite{bh10a}
\begin{equation}
\|\qm\|_{p}:=\Bigl(\Tr\bigl(\qm^{p}\bigr)\Bigr)^{1/p}
=\left(\sum\nolimits_{j=1}^{m}\lambda_{j}(\qm)^{p}{\,}\right)^{1/p}
\ . \label{schad}
\end{equation}
When $\qm\in\lpp(\hh)$, i.e. the $\qm$ is also invertible, the
right-hand side of (\ref{schad}) determines a symmetric anti-norm
with negative exponent $p<0$ \cite{bh10a,bh10b}. Using given
anti-norm (\ref{funex}), for any $p\in(0;1)$ we can construct
another anti-norm
\begin{equation}
\qm\mapsto\|\qm^{p}\|_{!}^{1/p}
\ . \label{afunex}
\end{equation}
This result posed in proposition 3.7 of \cite{bh10a} is based on
the following. Namely, the function $t\mapsto{t}^{p}$ with $p\in(0;1)$ is
matrix concave on positive matrices. Applying (\ref{afunex}) to (\ref{kyfad}) leads to
anti-norms
\begin{equation}
\|\qm\|_{\{k\}}^{(p)}:=\left(\|\qm^{p}\|_{\{k\}}\right)^{1/p}=
\Bigl\{\sum\nolimits_{j=1}^{k}\bigl(\lambda_{j}(\qm)^{\uparrow}\bigr)^{p}{\,}\Bigr\}^{1/p}
\ . \label{apgf}
\end{equation}
In a certain sense, the two-parametric family (\ref{apgf}) is an
anti-norm counterpart of the norm family (\ref{npgf}). Many
important results on matrix norms allow appropriate reformulations
with anti-norms \cite{bh10a,bh10b}. We shall now characterize
changes of the anti-norms (\ref{apgf}) under the operation of
partial trace.

\newtheorem{t51}[t31]{Proposition}
\begin{t51}\label{prop5}
Let $\wtq\in\lsp\bigl(\hh_{A}\otimes\hh_{B}\bigr)$,
${\rm{dim}}(\hh_{A})=m$, ${\rm{dim}}(\hh_{B})=n$, and
$\qm_{A}=\Tr_{B}\bigl(\wtq\bigr)$. For all $k=1,\ldots,m$ and
$p\in(0;1]$, there holds
\begin{equation}
\|\qm_{A}\|_{\{k\}}^{(p)}\geq{n}^{(p-1)/p}{\>}\|\wtq\|_{\{kn\}}^{(p)}
\ . \label{kqn1}
\end{equation}
For $\wtq\in\lpp\bigl(\hh_{A}\otimes\hh_{B}\bigr)$ and $p<0$,
the Schatten $p$-anti-norm satisfies
\begin{equation}
\|\qm_{A}\|_{p}\geq{n}^{(p-1)/p}{\>}\|\wtq\|_{p}
\ . \label{kqn2}
\end{equation}
\end{t51}

{\bf Proof.} For positive $\wtq$, each of $n^{2}$ summands in the
left-hand side of (\ref{xzrel}) is also positive. Using the
superadditivity inequality (\ref{sups}), the homogeneity, and the
symmetry of anti-norms, we then obtain
\begin{align}
\bigl\|\qm_{A}\otimes\pen_{B}\bigr\|_{\{kn\}}^{(p)}&\geq
\frac{1}{n}{\>}\sum_{l=0}^{n-1}\sum_{j=0}^{n-1}
\left\|\bigl(\pen_{A}\otimes\xm_{B}^{l}\zm_{B}^{j}\bigr){\,}
\wtq{\,}\bigl(\pen_{A}\otimes\xm_{B}^{l}\zm_{B}^{j}\bigl)^{\dagger}\right\|_{\{kn\}}^{(p)}
\nonumber\\
&=n{\,}\|\wtq\|_{\{kn\}}^{(p)}
\ . \label{kqn12}
\end{align}
Repeating the spectrum of $\qm_{A}$ by $n$ times, one then obtains
the spectrum of $\qm_{A}\otimes\pen_{B}$. So we can write
\begin{equation}
\bigl\|\qm_{A}\otimes\pen_{B}\bigr\|_{\{kn\}}^{(p)}=
\Bigl\{n\sum\nolimits_{j=1}^{k}\bigl(\lambda_{j}(\qm_{A})^{\uparrow}\bigr)^{p}{\,}\Bigr\}^{1/p}
=n^{1/p}{\,}\|\qm_{A}\|_{\{k\}}^{(p)}
\ . \label{kqn13}
\end{equation}
Combining (\ref{kqn12}) and (\ref{kqn13}) finally leads to
(\ref{kqn1}). For strictly positive $\wtq$, each summand in the
left-hand side of (\ref{xzrel}) is strictly positive as well. By a
parallel argument, for $p<0$ we have the relation
\begin{equation}
n^{1/p}{\,}\|\qm_{A}\|_{p}=\bigl\|\qm_{A}\otimes\pen_{B}\bigr\|_{p}\geq{n}{\,}\|\wtq\|_{p}
\ , \label{kq14}
\end{equation}
whence the claim (\ref{kqn2}) is provided. $\blacksquare$

The inequality (\ref{kqn1}) gives a lower bound on the anti-norms
(\ref{apgf}) of partial trace $\qm_{A}=\Tr_{B}\bigl(\wtq\bigr)$ in
terms of similar anti-norms of $\wtq$ and traced-out
dimensionality. This relation is sharp in the sense that it is
always saturated in the case (\ref{wrqa}) with any
$\rn_{A}\in\lsp(\hh_{A})$. Setting $k=m$ in (\ref{kqn1}), we
obtain the inequality (\ref{kqn2}) for $p\in(0;1]$ and positive
$\wtq$. The further choice $p=1$ actually gives the trace norm. It
must be stressed that the trace norm is justly an anti-norm on
positive matrices. Indeed, the superadditivity inequality
(\ref{sups}) is fulfilled here with equality. For the trace norm,
the relations (\ref{kpn1}) and (\ref{kqn1}) give
$\Tr(\qm_{A})\leq\Tr\bigl(\wtq\bigr)$ and
$\Tr(\qm_{A})\geq\Tr\bigl(\wtq\bigr)$, respectively. In fact, the
matrices $\qm_{A}=\Tr_{B}\bigl(\wtq\bigr)$ and $\wtq$ have the
same trace. Choosing $p=1$ in (\ref{kqn1}), we have the relation
in terms of the Ky Fan anti-norms, namely
\begin{equation}
\|\qm_{A}\|_{\{k\}}\geq\|\wtq\|_{\{kn\}}
\ . \label{kqk1}
\end{equation}
In view of $\Tr(\qm_{A})=\Tr\bigl(\wtq\bigr)$ and (\ref{kyfad}),
the inequality (\ref{kqk1}) is actually equivalent to the
inequality (\ref{kpk1}), in which the $k$ is replaced with
$(m-k)$.

Similarly to the case of norms, the relation (\ref{kqn1}) can be
combined with the Stinespring representation (\ref{strp}).
Obviously, we obtain
\begin{equation}
\|\phim(\qm)\|_{\{k\}}^{(p)}\geq{d}^{(p-1)/p}{\>}\bigl\|\wtv\qm\wtv^{\dagger}\bigr\|_{\{kd\}}^{(p)}
\ , \label{strp4}
\end{equation}
where $d={\rm{dim}}(\hh_{C})$. Consider trace-preserving
completely positive map $\phim$. With input $\qm\in\lsp(\hh_{A})$,
the right-hand side of (\ref{strp}) is partial trace of the
positive entry $\wtv\qm\wtv^{\dagger}$, in which $\wtv$ is an
isometry. Thus, the positive matrices $\wtv\qm\wtv^{\dagger}$ and
$\qm$ have the same non-zero eigenvalues, whence
\begin{equation}
\bigl\|\wtv\qm\wtv^{\dagger}\bigr\|_{\{kd\}}^{(p)}=\|\qm\|_{\{kd\}}^{(p)}
\ . \label{strp5}
\end{equation}
As
${\rm{dim}}\bigl(\hh_{B}\otimes\hh_{C})=nd\neq{m}={\rm{dim}}(\hh_{A})$
in general, we must stress a sense, in which the formula
(\ref{strp5}) does hold. The definition (\ref{apgf}) puts the
eigenvalues in non-decreasing order. If $nd>m$ then we supply the
spectrum of $\qm$ by $(nd-m)$ zeros and only afterward calculate
the right-hand side of (\ref{strp5}). With this convention, we
have arrived at a conclusion.

\newtheorem{ct2}[ct1]{Corollary}
\begin{ct2}\label{crl2}
Let $\phim:{\>}\lnp(\hh_{A})\rightarrow\lnp(\hh_{B})$ be
trace-preserving completely positive map with
${\rm{dim}}(\hh_{B})=n$, and let $\qm\in\lsp(\hh_{A})$. For all
$k=1,\ldots,n$ and $p\in(0;1]$, there holds
\begin{equation}
\|\phim(\qm)\|_{\{k\}}^{(p)}\geq{d}^{(p-1)/p}{\>}\|\qm\|_{\{kd\}}^{(p)}
\ , \label{stct2}
\end{equation}
where $d$ is the dimensionality of $\hh_{C}$ in the Stinespring
representation (\ref{strp}).
\end{ct2}

For any positive input $\qm$, the inequality (\ref{stct2}) gives a
lower bound on the anti-norms (\ref{apgf}) of the channel output
$\phim(\qm)$. In particular, the Schatten $p$-anti-norm obeys
\begin{equation}
\|\phim(\qm)\|_{p}\geq{d}^{(p-1)/p}{\>}\|\qm\|_{p}
\ , \label{stctpp}
\end{equation}
where real $p\in(0;1]$. We will use the above results for norms
and anti-norms in studying relations between the unified entropies
of a composite quantum system and one of its subsystems. We
conclude this section with an anti-norm counterpart of Proposition
\ref{prop2}. Such a result for anti-norms can be used in
combination with (\ref{kqn1}).

\newtheorem{t61}[t31]{Proposition}
\begin{t61}\label{prop6}
Let $\rn\in\lsp(\hh)$ and ${\rm{dim}}(\hh)=m$. For all
$k=1,\ldots,m$ and $p,q\in(0;1)$, there holds
\begin{equation}
\|\rn\|_{\{k\}}^{(p)}\geq{k}^{(q-1)/(pq)}{\>}\|\rn\|_{\{k\}}^{(pq)}
\ , \label{tpn62}
\end{equation}
with equality if and only if multiplicity of the smallest
eigenvalue of $\rn$ is not less than $k$.
\end{t61}

{\bf Proof.} One of the forms of H\"{o}lder's inequality states
the following (see, e.g., theorem 13 in \cite{hardy}). For two
$k$-tuples of positive numbers, we have
\begin{equation}
\sum\nolimits_{j=1}^{k}x_{j}{\,}y_{j}\geq
\Bigl(\sum\nolimits_{j=1}^{k}x_{j}^{q}\Bigr)^{1/q}
\Bigl(\sum\nolimits_{j=1}^{k}y_{j}^{r}\Bigr)^{1/r}
\ , \label{vhin13}
\end{equation}
where non-zero $q<1$ and $1/q+1/r=1$. Taking
$x_{j}=\bigl(\lambda_{j}(\rn)^{\uparrow}\bigr)^{p}$ and $y_{j}=1$,
we actually rewrite (\ref{gear}) in the opposite direction with
$\lambda_{j}(\rn)^{\uparrow}$ instead of
$\sigma_{j}(\rn)^{\downarrow}$. Raising the latter to the power
$1/p>0$, we herewith obtain (\ref{tpn62}). The inequality
(\ref{vhin13}) is saturated if and only if $x_{j}^{q}$ and
$y_{j}^{r}$ are proportional. Applying this point, we have arrived
at the above condition for equality. $\blacksquare$

\section{Inequalities for unified entropies}\label{inuen}

In this section, we derive some inequalities between the
$(\alpha,s)$-entropies of density operator
$\wro\in\lsp(\hh_{A}\otimes\hh_{B}\bigr)$ and one of its partial
traces $\bro_{A}=\Tr_{B}\bigl(\wro\bigr)$ and
$\bro_{B}=\Tr_{A}\bigl(\wro\bigr)$. We first recall the
definitions of used entropic measures. Let $\hh$ be
$m$-dimensional Hilbert space, i.e. ${\rm{dim}}(\hh)=m$. For
$\alpha>0\neq1$, the R\'{e}nyi entropy of density operator
$\bro\in\lsp(\hh)$ is defined as \cite{bengtsson}
\begin{equation}
\hr_{\alpha}(\bro):=\frac{1}{1-\alpha}{\ }\ln\bigl[\Tr\bigl(\bro^{\alpha}\bigr)\bigr]
=\frac{\alpha}{1-\alpha}{\ }\ln\|\bro\|_{\alpha}
\ . \label{qredf}
\end{equation}
Here the quantity
$\|\bro\|_{\alpha}=\bigl[\Tr(\bro^{\alpha})\bigr]^{1/\alpha}$ is
an anti-norm for $\alpha\in(0;1)$ and a norm for
$\alpha\in(1;\infty)$. The entropy (\ref{qredf}) is a quantum
counterpart of the classical entropy introduced by R\'{e}nyi
\cite{renyi61}. It reaches its maximal value $\ln{m}$ with the
completely mixed state $\bro_{*}=\pen/m$ on $\hh$. Another
extension of the standard entropy is the non-extensive entropy, or
the Tsallis entropy \cite{tsallis}. This concept is widely used in
much many topics of science \cite{AO01}. In quantum regime, the
non-extensive entropy of degree $\alpha>0\neq1$ is defined as
\begin{equation}
\rh_{\alpha}(\bro):=\frac{1}{1-\alpha}{\ }\Tr\bigl(\bro^{\alpha}-\bro\bigr)
=-\Tr\bigl(\bro^{\alpha}\ln_{\alpha}\bro\bigr)
\ , \label{qtadf}
\end{equation}
where $\ln_{\alpha}x=\bigl(x^{1-\alpha}-1\bigr)/(1-\alpha)$ is the
$\alpha$-logarithm. For the entropy (\ref{qtadf}), the maximal
value $\ln_{\alpha}m$ is also reached with $\bro_{*}=\pen/m$.
Bounds of Fannes type were derived for the Tsallis entropy itself
\cite{yanagi,zhang} and its partial sums \cite{rast1023}. Such
estimates are required in studying stability properties of various
entropies \cite{CL09}. The stability issue was inspired by Lesche
\cite{lesche}, who showed that the R\'{e}nyi $\alpha$-entropy is
not stable in the thermodynamic limit for all $\alpha>0\neq1$. In
the limit $\alpha\to1$, the definitions (\ref{qredf}) and
(\ref{qtadf}) both lead to the von Neumann entropy
${\mathrm{S}}(\bro)=-\Tr\bigl(\bro\ln\bro\bigr)$. General
references on the von Neumann entropy are the review \cite{wehrl}
and the comprehensive book \cite{ohya}. As is shown in
\cite{hey06}, the R\'{e}nyi and Tsallis entropies can both be
treated as particular cases of the following entropic functional.
For $\alpha>0\neq1$ and $s\neq0$, the unified $(\alpha,s)$-entropy
is defined by \cite{hey06}
\begin{equation}
\ers(\bro):=\frac{1}{(1-\alpha){\,}s}{\>}
\Bigl\{\bigl[\Tr(\bro^{\alpha})\bigr]^s-1\Bigr\}
=\frac{\|\bro\|_{\alpha}^{\alpha{s}}-1}{(1-\alpha){\,}s}
\ , \label{qundef}
\end{equation}
The formula (\ref{qundef}) gives the Tsallis entropy (\ref{qtadf})
for $s=1$ and the R\'{e}nyi entropy (\ref{qredf}) in the limit
$s\to0$. It is considered in \cite{rastjst} that the quantum
$(\alpha,s)$-entropy (\ref{qundef}) enjoys many properties
similarly to the von Neumann entropy. For instance, uniform
estimates were obtained in a wide parametric range \cite{rastjst}.
Entropies of the form (\ref{qundef}) have been used for a
unification of monogamy inequalities in multi-qubit systems
\cite{kims11}. In the finite-dimensional case, the
$(\alpha,s)$-entropy is bounded from above for $\alpha>0$ and all
real $s$, namely \cite{hey06}
\begin{equation}
\ers(\bro)\leq\ers(\bro_{*})=\frac{m^{(1-\alpha)s}-1}{(1-\alpha){\,}s}
\ . \label{maxeq}
\end{equation}
Entropies of probability distributions are obtained by replacing
the traces with the corresponding sums. In quantum theory, such
entropies are used for expressing uncertainties in measurements
\cite{ww10}. Formulations in terms of the unified entropies were
given in \cite{rast12qic}. In this section, we will deal only with
the quantum entropies. Our first result is posed as follows.

\newtheorem{t41}[t31]{Proposition}
\begin{t41}\label{prop3}
Let $\wro\in\lsp\bigl(\hh_{A}\otimes\hh_{B}\bigr)$ be a density
matrix, ${\rm{dim}}(\hh_{B})=n$, and
$\bro_{A}=\Tr_{B}\bigl(\wro\bigr)$. For all $\alpha>0\neq1$ and
$s\neq0$, there holds
\begin{equation}
\ers\bigl(\wro\bigr)\leq{n}^{(1-\alpha)s}{\>}\ers\bigl(\bro_{A}\bigr)
+\frac{1}{s}{\>}\ln_{\alpha}\bigl(n^{s}\bigr)
\ . \label{et41}
\end{equation}
\end{t41}

{\bf Proof.} Raising (\ref{kqn2}) and (\ref{spn1}), with $\alpha$
instead of $p$, to the power $\alpha$, we have arrived at the
relations
\begin{equation}
\|\wro\|_{\alpha}^{\alpha}
\left\{
\begin{array}{cc}
\leq, & \alpha\in(0;1) \\
\geq, & \alpha\in(1;\infty)
\end{array}
\right\}
n^{1-\alpha}{\,}\|\bro_{A}\|_{\alpha}^{\alpha}
\ . \label{pt401}
\end{equation}
For all $s\neq0$, the function
$t\mapsto(1-\alpha)^{-1}{s}^{-1}t^{s}$ is increasing for
$\alpha\in(0;1)$ and decreasing for $\alpha\in(1;\infty)$ (its
derivative is positive for the former and negative for the
latter). Applying this with (\ref{pt401}), we obtain
\begin{equation}
\frac{1}{(1-\alpha){\,}s}{\>}\|\wro\|_{\alpha}^{\alpha{s}}\leq
\frac{n^{(1-\alpha)s}}{(1-\alpha){\,}s}{\>}\|\bro_{A}\|_{\alpha}^{\alpha{s}}
\ . \label{pt411}
\end{equation}
Substituting this into the definition of $\ers\bigl(\wro\bigr)$
gives
\begin{equation}
\ers\bigl(\wro\bigr)\leq
\frac{1}{(1-\alpha){\,}s}{\>}\Bigl\{n^{(1-\alpha)s}\bigl(\|\bro_{A}\|_{\alpha}^{\alpha{s}}-1\bigr)+n^{(1-\alpha)s}-1\Bigr\}
\ . \label{pt412}
\end{equation}
The last expression is actually the right-hand side of
(\ref{et41}). $\blacksquare$

An important particular case of (\ref{et41}) takes place for
$s=1$. For all $\alpha>0$, the corresponding Tsallis
$\alpha$-entropies satisfy
\begin{equation}
\rh_{\alpha}\bigl(\wro\bigr)\leq{n}^{1-\alpha}{\>}\rh_{\alpha}\bigl(\bro_{A}\bigr)+\ln_{\alpha}n
\ . \label{ett41}
\end{equation}
Here the case of von Neumann entropy holds in the limit
$\alpha\to1$. The inequalities (\ref{et41}) and (\ref{ett41})
provide an upper bound on the entropies of joint density matrix in
terms of the reduced density and traced-out dimensionality. If
${\rm{dim}}(\hh_{A})=m$, then
${\rm{dim}}\bigl(\hh_{A}\otimes\hh_{B}\bigr)=mn$ and for all
$\alpha>0$ we have
\begin{align}
&\ers\bigl(\wro\bigr)\leq\frac{(mn)^{(1-\alpha)s}-1}{(1-\alpha){\,}s}
\ , \label{mxq}\\
&\rh_{\alpha}\bigl(\wro\bigr)\leq\ln_{\alpha}(mn)=n^{1-\alpha}\ln_{\alpha}m+\ln_{\alpha}n
\ . \label{mxqq}
\end{align}
When $\bro_{A}=\pen_{A}/m$, i.e. the state of subsystem $A$ is
completely mixed, the right-hand sides of (\ref{et41}) and
(\ref{ett41}) concur with the right-hand sides of (\ref{mxq}) and
(\ref{mxqq}), respectively. So the latter inequalities are covered
by the former as a very particular case. For the R\'{e}nyi
entropies, the following statement takes place.

\newtheorem{t42}[t31]{Proposition}
\begin{t42}\label{prop4}
Let $\wro\in\lsp\bigl(\hh_{A}\otimes\hh_{B}\bigr)$ be a density
matrix, ${\rm{dim}}(\hh_{B})=n$, and
$\bro_{A}=\Tr_{B}\bigl(\wro\bigr)$. For all $\alpha>0$, there
holds
\begin{equation}
\hr_{\alpha}\bigl(\wro\bigr)\leq\hr_{\alpha}\bigl(\bro_{A}\bigr)+\ln{n}
\ . \label{et42}
\end{equation}
\end{t42}

{\bf Proof.} Taking the logarithm of (\ref{pt401}), one obtains
\begin{equation}
\alpha\ln\|\wro\|_{\alpha}
\left\{
\begin{array}{cc}
\leq, & \alpha\in(0;1) \\
\geq, & \alpha\in(1;\infty)
\end{array}
\right\}
\alpha\ln\|\bro_{A}\|_{\alpha}+(1-\alpha)\ln{n}
\ . \label{pt413}
\end{equation}
We have $1-\alpha>0$ for $\alpha\in(0;1)$ and $1-\alpha<0$ for
$\alpha\in(1;\infty)$. Dividing (\ref{pt413}) by $(1-\alpha)$ then
provides (\ref{et42}) in view of (\ref{qredf}). The case of von
Neumann entropy is resolved in the limit $\alpha\to1$.
$\blacksquare$

The relation (\ref{et42}) is consistent with the relation
(\ref{et41}), in which the limit $s\to0$ is taken. We have
preferred a direct proof, since the case of R\'{e}nyi entropies is
especially important. For all $\alpha>0$, we can write
\begin{equation}
\hr_{\alpha}\bigl(\wro\bigr)\leq\ln(mn)=\ln{m}+\ln{n}
\ . \label{mxr}
\end{equation}
The right-hand side of (\ref{et42}) concurs with the right-hand
side of (\ref{mxr}), when $\bro_{A}=\pen_{A}/m$. So, the
inequality (\ref{mxr}) is a particular case of (\ref{et42}).

The presented entropic bounds are directly based on the relations
(\ref{spn1}) and (\ref{kqn2}). In a similar manner, the relations
(\ref{stctp}) and (\ref{stctpp}) lead to bounds with the unified
entropies of the output of a quantum channel. For trace-preserving
completely positive map $\phim$ and input density matrix $\bro$,
we have
\begin{equation}
\ers(\bro)\leq{d}^{(1-\alpha)s}{\>}\ers\bigl(\phim(\bro)\bigr)
+\frac{1}{s}{\>}\ln_{\alpha}\bigl(d^{s}\bigr)
\ , \label{stctep}
\end{equation}
including the bound
$\hr_{\alpha}(\bro)\leq\hr_{\alpha}\bigl(\phim(\bro)\bigr)+\ln{d}$
in the case $s=0$. Recall that $d$ is the dimensionality of
$\hh_{C}$ in (\ref{strp}) and also equal to the rank of the Choi
matrix. We can derive (\ref{stctep}) by obvious changes in the
proofs of this section. The bound (\ref{stctep}) reflects the fact
that $d={\rm{dim}}(\hh_{C})$ is a genuine characteristic of given
quantum channel.

The relations of this section are of general form without any
specifications. In some special cases, more detailed bounds could
be done. By one of the Lindblad inequalities \cite{glind91}, the
entropy exchange is bounded from above by the sum of the input and
output von Neumann entropies of the principal quantum system. The
entropy exchange can be posed as the output entropy of the
bipartite system composed of the principal and reference ones
\cite{nielsen}. Here the input bipartite state is pure and the
reference system itself is not altered. The Lindblad inequalities
have found use in studying additivity properties of quantum
channels \cite{rzf11}. Extensions of Lindblad's inequalities with
some of the unified entropies have been obtained \cite{rast11a}.
Applications of the bounds (\ref{et41}), (\ref{ett41}), and
(\ref{et42}) to entropic characteristics of quantum channels will
be discussed in a following work.

\begin{acknowledgments}
The author is grateful to anonymous referee for very useful
comments, including use of obtained results through the
Stinespring representation.
\end{acknowledgments}

\end{document}